\documentclass[twocolumn,amssymb,nobibnotes,superscriptaddress,aps,prl]{revtex4-2}
\usepackage{graphicx}
\usepackage{color,amssymb}
\usepackage{hyperref}
\usepackage{amsmath}

\begin{document}

\title{Collective charge excitations between moir\'e-minibands in twisted WSe$_2$ bilayers from resonant inelastic light scattering}%

\newcommand{\UM}{Institute of Physics, University of Münster, Wilhelm-Klemm-Str. 10, 48149 Münster, Germany}
\newcommand{\SoN}{Center for Soft Nanoscience (SoN), Busso-Peus-Str. 10, 48149 Münster, Germany}
\newcommand{\RWTH}{Institute for Theory of Statistical Physics, RWTH Aachen University, and JARA Fundamentals of Future Information Technology, 52062 Aachen, Germany}
\newcommand{\MPG}{Max Planck Institute for the Structure and Dynamics of Matter, Center for Free Electron Laser Science, 22761 Hamburg, Germany}
\newcommand{\UHH}{Institute of Theoretical Physics, University of Hamburg, Notkestrasse 9, 22607 Hamburg, Germany}
\newcommand{\HIC}{The Hamburg Centre for Ultrafast Imaging, 22761 Hamburg, Germany}

\author{Nihit Saigal}
    \affiliation{\UM}
    
\author{Lennart Klebl}
    \affiliation{\UHH}
    
\author{Hendrik Lambers}
    \affiliation{\UM}
    
\author{Sina Bahmanyar} 
   \affiliation{\UM}
   
\author{Veljko Anti\'{c}}
    \affiliation{\UM}
 
\author{Dante M. Kennes}
    \affiliation{\RWTH}
    \affiliation{\MPG}

\author{Tim O. Wehling}
    \affiliation{\UHH}
    \affiliation{\HIC}

\author{Ursula Wurstbauer}
    \affiliation{\UM}
    \affiliation{\SoN}
    \email{wurstbauer@uni-muenster.de}

\date{\today}%
\begin{abstract}

We establish low-temperature resonant inelastic light scattering (RILS) spectroscopy as a tool to probe the formation of a series of moir\'e-bands in twisted WSe$_{2}$ bilayers by accessing collective inter-moir\'e-band excitations (IMBE). We observe resonances in RILS spectra at energies in agreement with inter-moir\'e band transitions obtained from an \emph{ab-initio} based continuum model. Transitions between the first and second inter-moir\'e band for a twist angle of about $8^\circ$ are reported and between first and second, third and higher bands for a twist of about $3^\circ$. The signatures from IMBE for the latter highlight a strong departure from parabolic bands with flat minibands exhibiting very high density of states in accord with theory. These observations allow to quantify the transition energies 
at the K-point where the states relevant for correlation physics are hosted. 
\end{abstract}

\maketitle


Twisted van der Waals (vdW) bilayers present a unique condensed matter platform to realize and control electronic correlation effects \cite{kennes2021moire,Tang2020,wu2018hubbard}. The large scale superlattice created by the superposition of the two layers at a slight rotational mismatch defines a reciprocal mini-Brillouin zone with nearly dispersionless (flat) bands as long as the layers hybridize sufficiently. The drastically reduced kinetic energy results in a very high density of states (DOS) and even van Hove physics in those bands, driving electrons into the correlated regime
~\cite{balents_moire_review_2020, pantaleon_non_moire_2023}. In several graphene-based systems correlated and ordered electronic phases are experimentally well-established ~\cite{cao_unconventional_2018,cao_correlated_2018,hao_SC_trilayer_2021,park_SC_trilayer_2021,zhang_Gr_Twist_Multilayer_2022}.
 In transition metal dichalcogenide (TMDC) based vdW stacks the absence of complications like topological obstructions have facilitated high-level microscopic many-body studies from early on \cite{PhysRevLett.122.086402} and explained the emergence of ordered, insulating and also different flavors of correlated metallic states of matter \cite{PhysRevB.104.075150,Xian2021MoS}. Experimentally, the aforementioned correlation effects have been realized \cite{Wang2020WeSe,Ghiotto2021}, while superconductivity has remained elusive with currently one report of an (unclear and controversial) zero resistance state~\cite{Wang2020WeSe} even though there are multiple theoretical studies~\cite{Ryee23,DK_Klebl_competition_2023,wu_pair-density-wave_2023}.
 
\begin{figure*}[!htb]
        \center{\includegraphics[width=\textwidth]
        {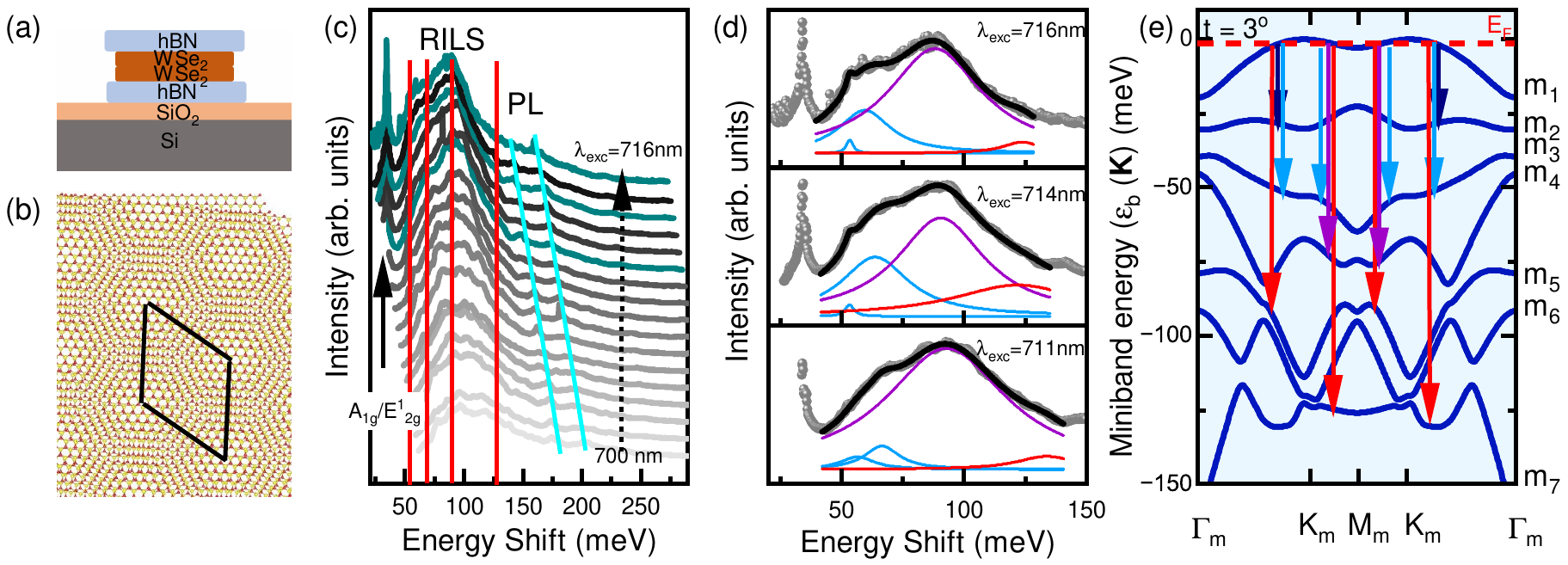}}
        \caption{\label{fig:Fig.1} 
(a) Scheme of the hBN encapsulated tWSe$_2$ bilayer.
(b) Stick-ball model of a moir\'e- superlattice from a tWSe$_2$ bilayer with the moir\'e unit cell indicated.
(c)  RILS spectra on a 3$^o$ tWSe$_2$ bilayer. Spectra are offset for clarity. The A$_{1g}$/E$^{1}_{2g}$ phonons are indicated by a black solid arrow. The broad RILS signal are collective electronic inter-moir\'e band excitations (IMBE). The vertical red solid lines mark resonance energies from theoretically predicted transition energies. The tilted cyan lines indicate emission signatures. For the dark cyan spectra results from line-shape analysis are shown in (d) [T = 4K, P$_{\rm Laser}$ = 1 mW].
(d) Selected RILS spectra (data points) from (c). The solid lines are fit results, using a set of Lorentzian lines to numerically deconvolute the individual contribution from the RILS spectra. The black lines are the sum of all Lorentzian curves.
(e)  Calculated energy dispersion of the 7 highest moir\'e minibands around the K, K' states for a 3$^o$ tWSe$_2$ bilayer. The minibands are labelled m$_1$, m$_2$, ...m$_7$. Exemplary inter-miniband transitions are sketched.
}
\end{figure*}

These emergent phases in semiconductor-based moir\'e bilayers are attributed to strongly interacting electronic K/K$^\prime$-states. In this letter, we experimentally access moir\'e minibands at the valence band maximum (VBM) around the K/K$^\prime$ valley by studying their collective electronic inter-moir\'e band excitations (IMBE) by resonant inelastic light scattering (RILS) experiments as summarized in Fig.~\ref{fig:Fig.1}. Accessing the moir\'e bands at the K-points is challenging for the combined reason of twist angle variations and reconstruction in realistic devices together with the VBM at the $\Gamma$ and K-points being close in energy \cite{Wang_2018}. The morphology of twisted bilayers, particularly at small twist angles, is such that variations in twist angle, but also reconstruction (in plane as well as corrugation) plays an important role. This leads to periodically patterned areas different to those expected from  a rigid moir\'e lattice-picture~\cite{Halbertal2021,andersen_excitons_2021,li_lattice_2021,zhao_excitons_2023}. While these patterns result in rich optical interband spectra \cite{zhao_excitons_2023} and even host coherent many-body states of excitons \cite{Troue23}, it makes the interpretation of spectroscopic signatures and the direct spectroscopy of moir\'e bands e.g. by angle-resolved photoemission spectroscopy (ARPES) difficult with only a few reports on selected materials combinations~\cite{karni22,Raja23}.
Recent $\mu-$ARPES and STM studies demonstrate the formation of moir\'e-bands at the valence band maximum (VBM) at the $\Gamma-$point of twisted WSe$_{2}$ bilayers \cite{zhang_flat_2020,li_lattice_2021,Gatti23}. Accessing moir\'e bands around the K/K' states and their properties is still lacking.\\

We remedy this standing problem by reporting on collective electronic excitations between moir\'e bands formed in twisted and hBN encapsulated WSe$_{2}$ bilayers [Fig.~\ref{fig:Fig.1}(a,b)] at the VBM around the K/K' states. By combined theoretical and experimental efforts, we demonstrate that probing the moir\'e-bands via collective single-particle-like IMBE by RILS spectroscopy analogue to intersubband excitations in 2D charge carrier systems hosted e.g. in GaAs quantum wells~\cite{Pinczuk71,Abstreiter1988,DasSarma99} provides a promising approach to study the band structure of twisted transition metal dichalcogenides at twist angles where correlation physics play an important role~\cite{wu2018hubbard,PhysRevB.104.075150,Xian2021MoS,Ryee23,DK_Klebl_competition_2023,wu_pair-density-wave_2023}.\\

The hBN encapsulated tWSe$_2$ bilayers have been prepared by micromechanical cleavage and viscoelastic dry transfer on top of Si/SiO$_2$ substrates with an estimated twist uncertainty of about $\pm0.5^o$. Three different types of WSe$_2$ samples have been prepared with a twist of ~3$^o$, ~8$^o$ and a natural homobilayer. To check for sufficient interlayer coupling and twist angle we employ low-temperature non-resonant Raman and PL spectroscopy (see SI-Figs.1, 2~\cite{SI}). For all measurements, the samples are mounted on the cold-finger of a closed-cycle refrigerator at a temperature of $T = 4$K. Position control is provided by x-y-z piezo actuators. The light from either a green solid state laser (2.33 eV) or a continuously tunable Ti:sapphire laser (linewidth of about 50 kHz) is focused with a cryogenic large-NA (NA = 0.82) objective lens to spot size of less than 2$\mu$m. The emitted/scattered light is guided to the entrance slit of a triple grating spectrometer. In RILS experiments, the sample is excited by linearly polarized light in back-scattering geometry and the scattered light is unpolarized (see SI-Fig.6 for polarization dependent spectra \cite{SI}). Due to the large NA of the objective a distribution of in-plane momenta $q_{\parallel} < 2\omega_{\rm L} / c \sin \theta_{\rm max}$ with $\theta_{\rm max} \thickapprox 55^o$ is transferred to the hole system. \\

By excitation in resonance close to the direct optical allowed interband transition at the K/K$^\prime$-point (A$_{1s}$-exciton), IMBEs at the K/K$^\prime$ VBM can be probed by low-temperature RILS spectroscopy.  Typical RILS spectra taken on a $t \thickapprox 3^\circ$ twisted WSe$_{2}$ bilayer at $T = 4$K are shown in a water fall representation in Fig.~\ref{fig:Fig.1}(c). The resonance excitation wavelengths are determined from PL experiments (see SI-Fig.1 \cite{SI}). A rich spectrum of rather wide and dispersive RILS modes (red lines) is acompagnied by the sharp nearly degenerate optical active A$_{1g}$, E$^{1}_{2g}$ phonon modes at an energy range between 20~mev and 120~meV. In order to deconvolute individual contributions to the RILS spectra, we perform a line-shape analysis by fitting a sum of Lorentzian curves to the spectra. We first focus on the energy range between 40~meV and 130~meV. The RILS spectra for all excitation wavelengths used are well reproduced by a sum of four (five) Lorentz functions for larger and smaller wavelengths, respectively. The energetic positions of the individual terms stay nearly constant, while the intensities are affected by the resonance conditions. This finding strongly suggests that the observed RILS mode originate from scattering on collective electronic excitation in analogy to intersubband excitations in GaAs based low-dimensional structures \cite{Pinczuk71, 1979_Abstreiter_SPE, Meier2021}. We would like to note that the quantitative interpretation from a similar quantitative line-shape analysis in the 20~meV to 35~meV is challenging since at least 7 phonon modes under the chosen resonance conditions are reported in literature \cite{McDonnell_2020}. Superimposed to these phonon modes and PL background, an additional mode occurs under extreme resonance at the red tail A$_{1g}$, E$^{1}_{2g}$ phonon modes at an energy of  about 28.8~meV  (for a detailed analysis see SI-Fig.5~\cite{SI}). Due to its occurrence only under extreme resonance in contrast to the resonantly activated phonon modes, we assign this mode also to an electronic excitation.

The extracted mode energies are in good quantitative agreement with inter-moir\'e band transitions extracted from the calculated electronic bands in the mini-BZ (mBZ) in the vicinity of the K/K$^\prime$ points. For direct comparison, Fig.~\ref{fig:Fig.1}(e) summarizes the first seven moir\'e minibands with examples of vertical transitions starting from the Fermi surface in the highest moir\'e band  m$_1$. We obtain the band structures of tWSe\textsubscript2 from a continuum model~\cite{wu2018hubbard, pan2020band, devakul2021magic, ryee2023switching} with parameters adjusted to \emph{ab-initio} simulations (for details see SI~\cite{SI}). 


\begin{figure*}[!htb] 
        \center{\includegraphics[width=1\textwidth]
        {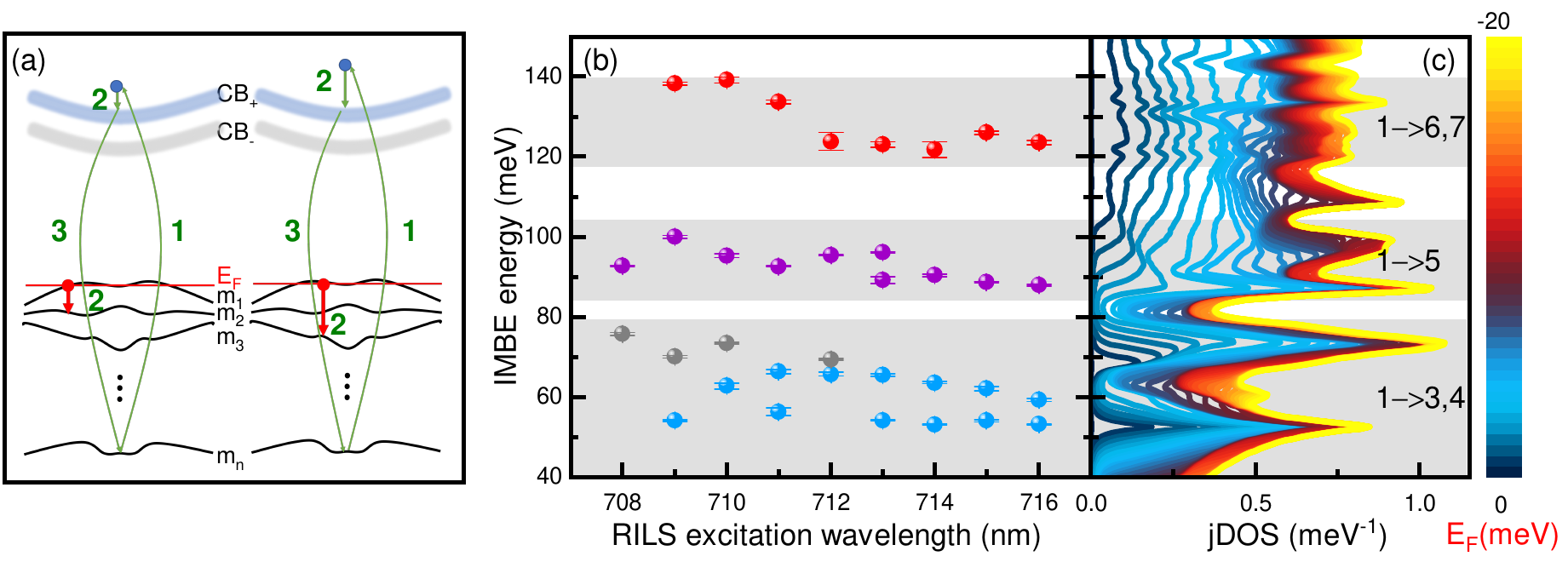}}
        \caption{\label{fig:Fig.2} 
 (a) Schematic picture of the three-step scattering process for creation of IMBE between moiré minibands formed in the VB around the K points. In the first step an electron is excited from a lower VB to a virtual state; in the second step the charger carrier is scattered by Coulomb interaction with photo-excited hole from $m_1$ to $ m_2$ (left) and from $m_1$ to $m_3$ (right) under creation of an IMBE. In the third step the electron recombines.
 (b) Extracted peak energies from the line-shape analysis described in Fig\ref{fig:Fig.1} for 3$^o$ tWSe$_2$. 
 (c) Calculated joint density of states jDOS for inter miniband transitions in dependence of the Fermi-level E$_F$ (color coded) of 3$^o$ tWSe$_2$ for vertical transitions. The peaks of the theoretical jDOS calculation are in good agreement with the Lorentzian fits to the experimental data (b).}
\end{figure*}

We explain the observed RILS spectra by a three-step scattering process for IMBE of photo-generated holes following the concept well-established for  single-particle intersubband excitations in GaAs-based films \cite{Pinczuk71, Abstreiter1988, DasSarma99} and successfully applied to e.g. RILS on intersubband excitation of photo-generated electrons and holes in ultrathin GaAs-based nanowires \cite{Meier2021}. Fig. \ref{fig:Fig.2}(a) illustrates the scattering processes for the $m_1 \rightarrow m_2$ (left) and $m_1 \rightarrow m_3$ (right) transitions: In the first step an electron from a VB state at K/K$^\prime$ is excited to a virtual state close to the upper CB band at K/K$^\prime$ (spin allowed transition). If the virtual electronic state is just above the CB by the energy of an IMBE, in the second step, the electron can be resonantly scattered via Coulomb interaction with a hole in $m_1$ miniband at the (quasi) Fermi-level to the CB. Simultaneously, an IMBE (e.g. $m_1 \rightarrow m_2$) is created in the valence band. In the third step, the scattered electron-hole pair recombines under emission of the the scattered photon. The same concept applies by including excitonic effects (not included in the scheme for clarity). The experiment is done under cw excitation such that a plasma of photo-generated holes is forming a quasi Fermi-level in the topmost miniband due to quick relaxation. We assume a charge imbalance between electrons and holes at the K/K$^\prime$ valley since the WSe$_2$ is expected to be slightly hole-doped and in addition photo-generated electrons are supposed to relax into the energetically lower $\Sigma$-valley \cite{2020_Brem_tWSe2}.\\

The hBN encapsulated tWSe$_2$ bilayers have been prepared by micromechanical cleavage and viscoelastic dry transfer on top of Si/SiO$_2$ substrates with an estimated twist uncertainty of about $\pm0.5^o$. Three different types of WSe$_2$ samples have been prepared with a twist of ~3$^o$, ~8$^o$ and a natural homobilayer. To check for sufficient interlayer coupling and twist angle we employ low-temperature non-resonant Raman and PL spectroscopy (see SI-Figs.1, 2~\cite{SI}). For all measurements, the samples are mounted on the cold-finger of a closed-cycle refrigerator at a temperature of $T = 4$K. Position control is provided by x-y-z piezo actuators. The light from either a green solid state laser (2.33 eV) or a continuously tunable Ti:sapphire laser (linewidth of about 50 kHz) is focused with a cryogenic large-NA (NA = 0.82) objective lens to spot size of less than 2$\mu$m. The emitted/scattered light is guided to the entrance slit of a triple grating spectrometer. In RILS experiments, the sample is excited by linearly polarized light in back-scattering geometry and the scattered light is unpolarized (see SI-Fig.6 for polarization dependent spectra \cite{SI}). Due to the large NA of the objective a distribution of in-plane momenta $q_{\parallel} < 2\omega_{\rm L} / c \sin \theta_{\rm max}$ with $\theta_{\rm max} \thickapprox 55^o$ is transferred to the hole system. 


\begin{figure*}[!htb]
        \center{\includegraphics[width=\textwidth]
        {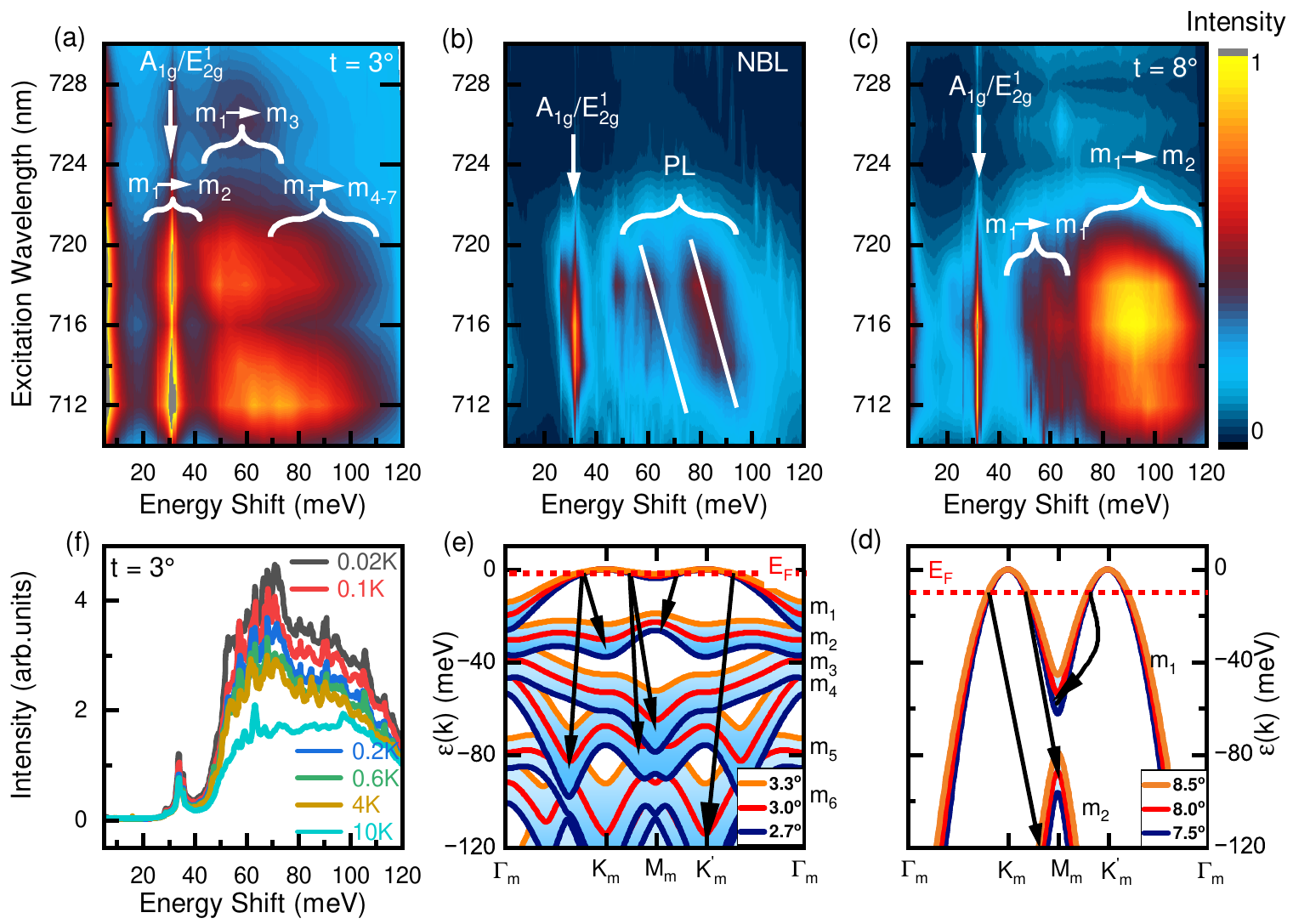}} 
        \caption{\label{fig:fig.3} 
RILS spectra on (a) 3$^o$ tWSe$_2$, (b) natural bilayer (NBL) and (c) WSe$_2$ 8$^o$ tWSe$_2$. The straight white arrows indicate phonon modes and the
white curly braces in (a) and (c) IMBE with moir\'e band assignment according to calculations. In the natural bilayer spectra (b) in addition to the vertical phonon line only shifting emission signals marked with tilted white lines emerge. In (c) an individual spectrum is exemplarily overlaid in grey color with a fit to the data (blue line) using the sum of two Lorentz curves.  [T = 4K, P$_{\rm laser}$ = 500$\mu$W; inset: $\lambda_{\rm laser}$=712nm].  
(d) Calculated electronic band structure for (8$^o \pm 0.5)^o$ considering realistic twist variations. 
(e) Calculated electronic band structure for (3$^o \pm 0.3)^o$. Twist variations can result in broadened minibands causing significantly broadened IMBE in RILS experiments. Breakdown of wave-vector conservation together with finite transferred in-plane momentum $q_\parallel$ due to the high NA of the objective allows also for non-vertical to points of the minibands with high DOS as indicated by the arrows. 
(f) Temperature dependent RILS spectra for 10mK $<$ T $<$ 10K for 3$^o$ tWSe$_2$. [$\lambda_{exc}$ = 710nm, P$_{\rm Laser}$ = 50 $\mu$W]. RILS signals fade out with increasing temperature and at 10K only a weak signal superimposed by emission background remains. 
}
\end{figure*}

Since jDOS is dependent on E$_F$, complementary excitation-power dependent RILS measurements are done in order to vary E$_F$ by the photo-induced holes from $\Delta$p $\thickapprox10^{9}$cm$^{-2}$ to $\Delta$p $\thickapprox10^{10}$cm$^{-2}$ and $\Delta$p $\thickapprox10^{12}$cm$^{-2}$. Due to the strong band non-parabolicity and the unknown intrinsic doping it was not possible to estimate the change in E$_F$ from $\Delta$p. In agreement with expectation from jDOS, the RILS intensities in the lower-energy region assigned to superimposed phonons and collective electronic excitations including the tentative  IMBE m$_1 \rightarrow$m$_2$ is reduced, while the modes are enhanced with slight modification in their energies in line with expectations from theory (spectra and summary from line-shape analysis shown in SI-Figs.3,4~\cite{SI}).

In Fig.\ref{fig:Fig.2}(b) we summarize IMBE-energies extracted from the lineshape analysis of the RILS spectra of the ~3$^o$ tWSe$_2$ described above. The nominal numerical fit error is only a few percent. The partially ascending slopes for few data points in dependence of the excitation wavelength is assigned to superimposed PL. The mode energies are directly compared to the calculated joint density of states (jDOS) of the individual transitions displayed in Fig.~\ref{fig:Fig.2}(c) in dependence of the position of E$_F$ (color code) demonstrating its significant E$_F$-dependency. Mode energies and the jDOS are in good quantitative agreement assuming a realistic Fermi-energy of less than a few meV. In the energy range between 50~meV and 80~meV two well separated modes can be identified that are assigned to two m$_1 \rightarrow$ m$_3$ and one m$_1 \rightarrow$ m$_4$ transitions, the third one might be another m$_1 \rightarrow$ m$_3$ superimposed by PL. The mode observed at an energy of about 95~meV is assigned to m$_1 \rightarrow$ m$_5$ transition and the modes observed around 123~meV and 140~meV to m$_1 \rightarrow$ m$_6$ and m$_1 \rightarrow$ m$_7$ and higher transitions, respectively. Minor deviation between theory and experiment can have following concurrent origins: (i) The moir\'e bands are flat but still dispersing in a highly non-parabolic manner such that they themselves have a finite width of up to few tens of meV with overlapping bands particularly for m$_3$ and higher. (ii) Non-vertical transitions between points with a high DOS are expected to contribute to finite momentum transfer $q_{\parallel}$ in addition to defect induced breakdown of momentum conservation \cite{Wurstbauer11}. (iii) Twist angle variations of at least $\pm 0.3^o$ within the laser spot has a crucial impact on the moir\'e bands formation as discussed below.\\ 
In Fig.~\ref{fig:fig.3}, we contrast RILS spectra in a false color representation taken on 3$^o$ tWSe$_2$ (a), natural bilayer (NBL) WSe$_2$ (b) and 8$^o$ tWSe$_2$ (c) for an extended range of excitation wavelengths. RILS intensities are encoded in the color scheme.
In the RILS spectra of the twisted bilayers 3$^o$ tWSe$_2$ and 8$^o$ tWSe$_2$ displayed in Figs.~\ref{fig:fig.3}(a) and (c), highly resonant, intense and rather broad RILS modes interpreted as IMBE dominate the RILS spectra. In addition to the phonon and IMBE resonances, a few features occur with energies depending on the incoming laser energy (tiled cyan lines) that are interpreted as defect or 0D-moir\'e potential localized emission signatures. The double resonance behaviour with a splitting of ~15~meV only observed for 3$^o$ is in agreement with reported splitting in absorption and emission spectra for lower twist angles \cite{andersen_excitons_2021}.

For the NBL sample despite the sharp phonon resonance, no clear RILS modes occur. The spectra are dominated by typical emission signatures. These lines are likely due to emission from interlayer excitons~\cite{Wang_2018}. The absence of RILS modes for NBL is expected due to the absence of moir\'e lattice and minibands. The energy separation of the SOI-split valence bands is in the order of a few hundreds of meV \cite{Kormanyos_2015} and hence far beyond the investigated energy range. For the 8$^o$ tWSe$_2$, instead, one weaker and one dominant broad RILS resonance appear at an energy range of about 60~meV to 120~meV. In this energy range, a good fit to the RILS spectra is achieved by a sum of two-Lorentzian terms [see example spectrum with fit overlaid in Fig. \ref{fig:fig.3}(c)]. For a quantitative comparison with theory, the related band structure for the moir\'e mBZ around the VBM at the K-point is displayed in Fig.~\ref{fig:fig.3} (d) in the experimentally relevant energy range (for extended range see SI-Fig.7~\cite{SI}). For such a large twist, the bands are nearly parabolic with larger energy separation. In comparison with theory, the weaker RILS mode at $\thickapprox$~60meV could be interpreted as a collective intraband transition within the m$_1$-band as sketched in Fig. \ref{fig:fig.3}(d). The more intense, broader RILS mode at $\thickapprox 75 - 115$~meV is interpreted as an IMBE m$_1\rightarrow$m$_2$ transition with the broadening due to non-vertical transitions provided by the finite transferred in-plane momentum $q_{\parallel} \leq q_{\rm max}$. The good quantitative agreement justifies our interpretation of the modes as being collective intra- and interband excitations allowing to study the band-structure of the mBZ at the K-points. In the following, we consider the impact of realistic twist-angle variations to be in the order of $\pm 0.3 ^o$. While it is shown from calculated bands to be negligible small for $(8\pm 0.5)^o$ tWSe$_2$, the impact is significant in the miniband energies and dispersion for $(3\pm 0.3)^o$ tWSe$_2$ as plotted in Fig.~\ref{fig:fig.3}(e). As highlighted above, twist angle variation together with non-vertical transitions and finite momentum transfer results in broadened IMBE signatures in RILS. The RILS modes are a convolution of jDOS and the momentum distribution. We would like to note that we do not observe sizeable differences for linear-co and linear-cross polarization in the RILS spectra (see SI-Fig.6~\cite{SI}). This finding supports our interpretation that the RILS mode are single-particle type of IMBE rather than plasma-like charge-density or spin-density excitations that show both a distinct polarization dependence \cite{Meier2021}. To further substantiate our interpretation, we perform temperature dependent RILS measurements (Fig. \ref{fig:fig.3}(f)). To avoid heating effects, the excitation power is reduced to P = 50$\mu$W and the excitation wavelength kept constant at extreme resonance condition at $\lambda$=710nm. The RILS signal shows a clear temperature dependence and is reduced with increasing temperature. Already at 10K the RILS modes fade out and the PL background starts to dominate the spectrum. We observe similar reduction in RILS intensity with increasing temperature for the $8^o$ tWSe$_2$ sample (see SI-Fig.6~\cite{SI}). All experimental observations together strongly support the interpretation that the observed broad RILS signatures are indeed collective IMBE between moir\'e bands at the VBM at the K/K$^\prime$ valley in excellent agreement to theory. We would like to emphasize that collective excitations probe the weakly disordered parts of the sample, while disorder often results in the most dominant signatures in emission experiments due to rather bright localized emission and disturbs transport investigations by additional scattering channels.


To conclude, we establish that low-temperature RILS experiments on collective "single particle like" \cite{Pinczuk71, DasSarma99, Meier2021} IMBE on twisted WSe$_2$ bilayers is a powerful method to experimentally study the moir\'e band formation selectively at the VBM around the K/K$^\prime$ states, their energetic separation and twist angle dependence providing a promising approach to study the band structure of twisted TMDCs at twist angles where correlation physics play an important role~\cite{wu2018hubbard,PhysRevB.104.075150,Xian2021MoS,Ryee23,DK_Klebl_competition_2023,wu_pair-density-wave_2023}. In agreement with theory, we identify several IMBE for a 3$^o$ tWSe$_2$, while for a 8$^o$ tWSe$_2$ only one clear IMBE and presumably one collective intraband excitation is observable and none natural bilayers. The observation of collective IMBE by RILS in a semiconducting vdW bilayer demonstrates the potential to access the collective low-lying excitation spectra as unique fingerprints of individual quantum phases by RILS similar to correlated phases e.g. in the fractional quantum Hall effect regime \cite{Wurstbauer13,Du_2019, Liang.2024}. 

\begin{acknowledgments}
The authors gratefully acknowledge the German Science Foundation (DFG) for financial support via Grants WU 637/7-1, WE 5342/5-1 and the Priority Program SPP 2244 ``2DMP'' - 443273985, 443274199 as well as the computing time granted through JARA on
the supercomputer JURECA~\cite{JUWELS} at Forschungszentrum Jülich. LK acknowledges support from the DFG through FOR 5249 (QUAST, Project No.
449872909). TW acknowledges support by the Cluster of Excellence `CUI: Advanced Imaging of Matter' of the DFG (EXC 2056, Project ID 390715994). DMK acknowledges support by the Deutsche Forschungsgemeinschaft (DFG, German Research Foundation) under Germany's Excellence Strategy - Cluster of Excellence Matter and Light for Quantum Computing (ML4Q) EXC 2004/1 - 390534769. We acknowledge support by the Max Planck-New York City Center for Nonequilibrium Quantum Phenomena.

\end{acknowledgments}


%


\newcommand{\bvec}[1]{\boldsymbol{#1}}

 \clearpage

\newpage
\pagebreak

\widetext
\begin{center}
\textbf{\large Supplemental Materials:\\
Collective charge excitations between moir\'e-minibands in twisted WSe$_2$ bilayers from resonant intelatic light scattering}
\end{center}

\setcounter{equation}{0}
\setcounter{figure}{0}
\setcounter{table}{0}
\setcounter{page}{1}
\makeatletter
\renewcommand{\theequation}{S\arabic{equation}}
\renewcommand{\thefigure}{S\arabic{figure}}
\renewcommand{\bibnumfmt}[1]{[S#1]}
\renewcommand{\citenumfont}[1]{S#1}


\section{Methods and experimental procedure}

\subsection{Sample preparation}

The tWSe$_2$ samples were prepared by first mechanically exfoliating the monolayer WSe$_2$ and multilayer hBN crystals onto viscoelastic PDMS stamps followed by their deterministic dry transfer (DDT) using a home-built setup with x-y-z and rotational micro-manipulators. To fix the stacking (twist) angle between the WSe$_2$ monolayers, their crystalline edges were determined from optical images on the PDMS stamps.  Following this, the top WSe$_2$ monolayer was stacked on top of the bottom WSe$_2$ monolayer at the desired twist angle by using a rotation stage attached to the sample stage in the DDT setup. After preparation of the full heterostructure including the hBN encapsulation layers, the samples were annealed in a home-built vacuum annealing chamber at a nominal pressure in the order of ~5x10$^{-5}$mbar at 100 - 120 $^o$C for 5 - 6 hours. An optical micrograph of a prepared 3$^o$ tWSe$_2$ sample as an example is shown in the inset of SI-Fig.1. 

\subsection{Photoluminescence (PL) and non-resonant Raman spectroscopy}

After preparation of the samples, they were characterized by low temperature non-resonant PL and Raman spectroscopies. The samples were mounted on the cold finger equipped with x-y-z piezo stages of a dilution refrigerator. The samples were  cooled down to 4K  using  pulse tube refrigeration operating with helium gas. For non-resonant characterization measurements, a 532 nm (2.33 eV) solid state laser, cleaned up using a grating based monochromator with a 1nm bandpass, was used for excitation of the samples. The laser was focused on the samples with a low temperature compatible objective lens with an NA of 0.82 mounted on a homebuilt stage attached to the 4K cryoshield of the refrigerator. The focused spot on the sample had a diameter of $\sim$ 2$\mu$m. The emission and scattering signals from the sample were collected by the same objective lens in a back scattering geometry and dispersed using a single spectrometer with a 600 l/mm grating for PL and an 1800 l/mm grating for Raman spectroscopy. The focal length of the spectrometer was 750 mm and the spectral resolution (532 nm, 1800 l/mm grating) was $\sim$ 0.4 cm$^{-1}$. The Rayleigh scattered light was filtered out by using a 532 nm long pass filter in front of the spectrometer. The dispersed light was collected and measured using a liquid nitrogen cooled CCD camera. Position and laser power dependent PL and position dependent Raman measurements were used to select the region of the sample for subsequent resonant inelastic light scattering measurements.

\subsection{Resonant Inelastic Light Scattering spectroscopy}

Following the characterization by low temperature PL and Raman spectroscopy, inelastic light scattering spectra under resonant excitation conditions were measured. The samples were mounted inside a dilution refrigerator as described above. The samples were excited at energies between 1.698 eV (730 nm) and 1.771 eV (700 nm) using a continuously frequency tunable Ti:sapphire laser stabilized by a ring cavity with a linewidth of about 50kHz. The luminescence background from the Ti:sapphire crystal was removed by passing the excitation laser beam through a monochromator (1nm bandpass). The light emitted or scattered from the sample surface were again collected in back scattering geometry and sent to either a single spectrometer with a 600 l/mm grating or to a triple spectrometer with a grating combination of 300 l/mm, 300 l/mm and 600 l/mm. The triple spectrometer was used in a subtractive mode where the second stage of the spectrometer is used to suppress the dispersion generated by the first stage by appropriate alignment of the gratings. This meant that at the exit slit of the second stage, light could be filtered out based on the angle of entrance in the spectrometer. Thus, the first two stages served as a bandpass filter and filtered out any stray light not following the path of light scattered from the sample, as well as the Rayleigh scattered background. 

\subsection{Determination of photo-excited hole density in power-dependent measurements}

For the estimation of the change in the photo-generated charge carrier density by excitation power dependent measurements, we start considering the generation rate of photo-generated electron-hole pairs given by:
\begin{equation}
    \label{eq:generation-rate}
    G = \frac{P \sigma}{A h \nu},
\end{equation}
where $P$ is the power of the incident light in units Watt, $\sigma$ is the absorbance of the sample at the specific excitation wavelength, $A$ is the area of the sample in unit cm$^{-2}$ and $h \nu$ is the energy of the photon in unit Joule. In the current experiment the excitation is done in continuous wave (cw) mode such that we can assume steady state conditions. The change in the photo-generated charge carries constitutes
\begin{equation}
    \frac{dp}{dt} = G-\frac{p}{\tau}=0
\end{equation}
and thus,
\begin{equation}
   p = G \tau = \frac{P \sigma \tau}{A h \nu}
\end{equation}
with $\tau$  the lifetime of the photo-excited charge carriers reported to be $\approx$100ps - 1ns \cite{2020_Scuri_WSe2-lifetime}. We approximate the spot area of the sample by $A \approx 6µ$m$^2$. The optical absorbance of the WSe$_2$/MoSe$_2$ heterostructure in the relevant range of excitation wavelength of 700nm to 716nm (1.73 - 1.77 eV) is taken from spectroscopic imaging ellipsometry investigations \cite{2022_Sigger_SIE}. We would like to note that the carrier generation rate, the absorbance as well as lifetime can be subject of the change of carrier density itself.
Following this approach, we approximate the change of the photogenerated charge carrier density, i.e. change in the hole density $\Delta p$ in dependence of the excitation power as following:

\begin{itemize}
    \item  P$_{Laser}$ = 1$\mu$W $\rightarrow \Delta$p  $\thickapprox10^9$cm$^{-2}$, 
    \item P$_{Laser}$ = 50$\mu$W $\rightarrow \Delta$p $\thickapprox10^{10}$cm$^{-2}$,
     \item P$_{Laser}$ = 1mW $\rightarrow \Delta$p $\thickapprox10^{12}$cm$^{-2}$.
\end{itemize}

We would like to note that the as exfoliated WSe$_2$ is typically slightly p-type doped. We assume the photo-generated holes for monolayers and also for the tWSe$_2$ bilayer to reside at the K, K\' valley forming an two-dimensional hole gas with a quasi-Fermi level under steady state conditions. Particularly for the bilayer the situation is expected to be different for the conduction band. The $\Sigma$-valley is energetically lower compared to the K-valley such that efficient charge carrier relaxation to the $\Sigma$-valley is expected \cite{2023_Bange}. As a direct consequence, an electron gas is expected to be formed rather at the $\Sigma$-valley compared to the K-valley. Therefore, at the K-valley probed in the RILS experiment in this study by the selected resonance conditions, an excess of holes is assumed such that the RILS experiment is sensitive to the collective excitation of the two-dimensional hole plasma formed at the VB top at the K valley around which the moir\'e minibands are formed.


\section{Theoretical calculations}
\subsection{Continuum Hamiltonian}
In the small-angle approximation $\Theta\sim0^\circ$, the dispersion of tWSe\textsubscript2 can be obtained by constructing a continuum model (following Refs.~\cite{wu2018hubbard, pan2020band, devakul2021magic, ryee2023switching}). As the electrons in a WSe\textsubscript2 monolayer are spin-valley locked, only one valley per layer contributes to each spin sector of the moir\'e Hamiltonian
\begin{equation}
    \label{eq:hamiltonian}
    H_\uparrow(\bvec k)= \begin{pmatrix}
        -\frac{\hbar^2(\bvec k - K_1)^2}{2m^\star} + \Delta_1(\bvec r) & \Delta_T(\bvec r) \\
        \Delta_T^\dagger(\bvec r) & -\frac{\hbar^2(\bvec k - K_2)^2}{2m^\star} + \Delta_2(\bvec r)
    \end{pmatrix} \,,
\end{equation}
where $m^\star = 0.43\,m_e$ is the effective mass, $K_l$ the original $K$-point of layer $l=1,2$, $\Delta_T(\bvec{r})$ the interlayer potential, and $\Delta_l(\bvec{r})$ the intralayer potential. The other spin sector $H_\downarrow(\bvec k)$ is obtained from time reversal symmetry. The inter- and intralayer potentials are parameterized by the following simple formulæ:
\begin{equation}
    \label{eq:potentials}
    \Delta_l(\bvec r) = 2V\,\sum_{\bvec g\in\triangle} \cos\big(\bvec g\cdot \bvec r + \sigma_z^{ll} \psi \big) \,, \qquad
    \Delta_T(\bvec r) = w\,\sum_{\bvec g\in\triangle} e^{-i\bvec g\cdot \bvec r} \,.
\end{equation}
Here, $\sigma_z$ is the Pauli $z$ matrix and $\bvec g$ are three reciprocal moir\'e lattice vectors that are connected by $C_3$ rotations. Hence we index the summation over those three reciprocal moir\'e lattice vectors as $\bvec g\in\triangle$. The $K$- and $K^\prime$-points in the moir\'e Brillouin zone (BZ) coincide with $K_1$ and $K_2$, respectively.

The intralayer potential strength $V=9\,\mathrm{meV}$, the interlayer hopping $w=18\,\mathrm{meV}$, and the intralayer potential parameter $\psi=128^\circ$ are taken to match density functional theory results~\cite{devakul2021magic}. Note that the twist angle $\Theta$ enters Eq.~(\ref{eq:hamiltonian}) through geometry in the momentum argument $\bvec k$ as well as the length of the inverse moir\'e lattice vectors $\bvec g$ in Eq.~(\ref{eq:potentials}).

For numerical diagonalization of the Hamiltonian Eq.~(\ref{eq:hamiltonian}), the expansion in inverse moir\'e lattice vectors must be truncated. We take into account the following inverse moir\'e lattice vectors $\bvec G$ 
\begin{equation}
    \bvec G \in \{ \bvec H ~|~ \| \bvec H \| \leq 3\| \bvec g \| ~\forall~ \bvec g \in \triangle \}\,,
\end{equation}
i.e., a distance shell cutoff with length $3$, which is sufficient for convergence of the resulting band structures.

\subsection{Joint Density of States}
The joint density of states (jDOS) data in Fig.~3~(f) of the main text is obtained numerically from the model discussed above. In essence, we implement the following formula:
\begin{equation}
    \label{eq:jdos}
    \mathrm{jDOS}(\omega) = \mathrm{Im} \, \frac1{N_{\bvec k}}\sum_{\bvec k, b_1, b_2} \frac{
        f\big(\epsilon_{b_1}(\bvec k)\big) - f\big(\epsilon_{b_2}(\bvec k)\big)
        }{\omega -i\eta + \epsilon_{b_1}(\bvec k) - \epsilon_{b_2}(\bvec k)} \,,
\end{equation}
where $\epsilon_b(\bvec k)$ is the dispersion of band $b$ with $\bvec k$ in the mini BZ, $f(x)=\big(1+\exp[\beta(x-\mu)]\big)^{-1}$ the Fermi function, and $\eta=5\,\mathrm{meV}$ a Lorentzian broadening. The inverse temperature is set to $\beta=10^2\,\mathrm{meV^{-1}}$. As we perform massive scans in frequency $\omega$ and chemical potential $\mu$ (Fermi energy) for significant $\bvec k$-meshes ($96\times96$ regularly spaced points in the primitive zone), we make use of the parallel nature of Eq.~(\ref{eq:jdos}) using a custom GPU implementation.

\clearpage
\section{Additional Figures} 

\subsection{Low-temperature photoluminescence spectrum of tWSe$_2$ ($t=3°$)}

\begin{figure}[!htb] 
       \centering
       \includegraphics[clip,width=0.5\columnwidth]
        {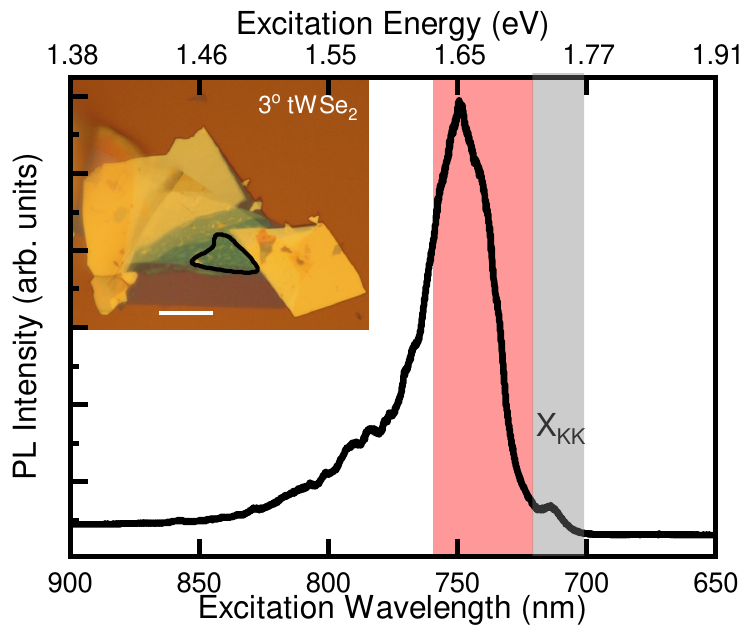}
        \caption{\label{fig:Fig.2} 
 (a) PL spectrum shows weak emission from direct K/K$^\prime$ excitons (X$_{KK}$) and a intense broad emission band from dark, charged and defect bound excitons. The spectral range around (X$_{KK}$) utilized for RILS spectroscopy is grey shaded and the region where modes in RILS are expected red shaded  [T = 4K, P$_{\rm Laser}$ = 1mW, E$_{\rm Laser}$ = 2.33eV]. Inset: Optical micrograph of a hBN encapsulated 3$^o$ tWSe$_2$ sample with active region marked marked by black line. [scale bar 20$\mu$m]
 }
\end{figure}
\clearpage

\subsection*{Non-resonant Raman measurements}

\begin{figure*}[!htb]
        \center{\includegraphics[width=0.6\textwidth]
        {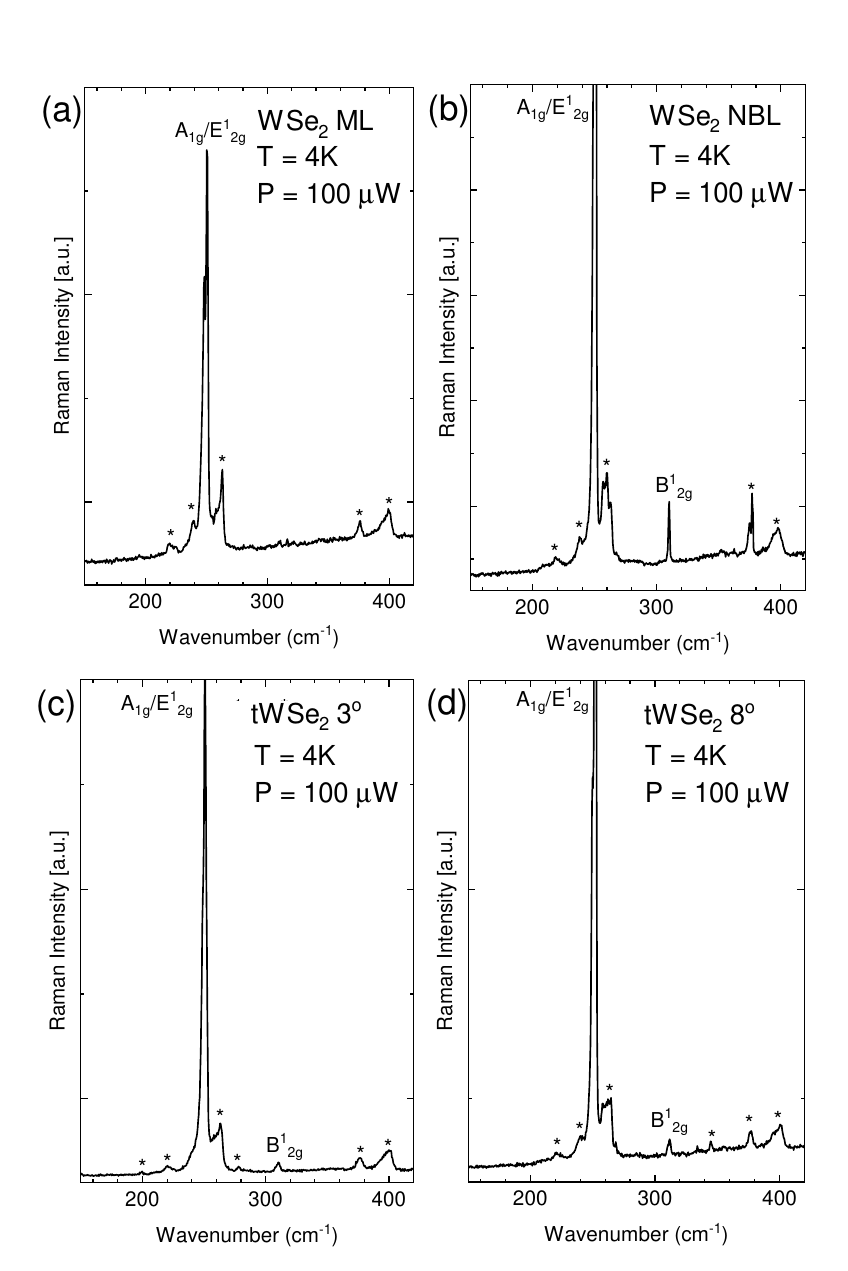}}
        \caption{\label{fig:NR_Raman} 
  (a) Non-resonant Raman spectra on hBN encapsulated WSe$_2$ monolayer, (b) hBN encapsulated natural bilayer (NBL) WSe$_2$, (c) hBN encapsulated  3$^o$ tWSe$_2$ and (c) hBN encapsulated 8$^o$ tWSe$_2$. All spectra show the optical phonon modes A$_{1g}$, E$^{1}_{2g}$ and in addition the NBL and twisted bilayers show the B$^{1}_{2g}$ phonon mode.  The features marked by an asterisk (*) are due to multiphonon scattering processes [T = 4K, P$_{laser}$ = 100$\mu$W, E$_{laser}$ = 2.33eV].}
\end{figure*}

\clearpage
\subsection*{Excitation power dependent RILS measurements on tWSe$_2$ ($t=3°$)}

\begin{figure*}[!htb]
        \center{\includegraphics[width=\textwidth]
        {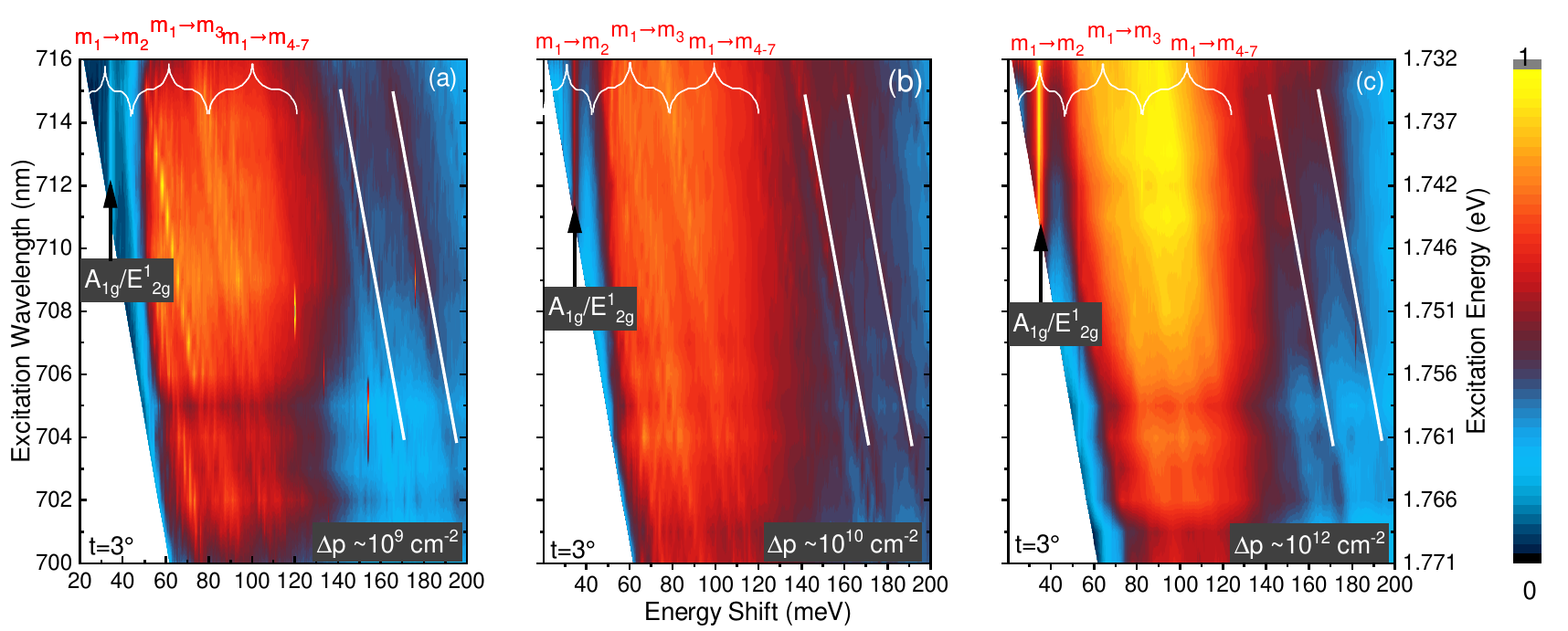}}
        \caption{\label{fig:Fig.SI_RILS_P} 
 (a) False color plots of RILS spectra for 3$^o$ tWSe$_2$ at three different excitation powers of (a) 1$\mu$W, (b) 50$\mu$W and (c) 1mW, respectively. The corresponding changes in the photo-excited hole-densities are $\sim{10^{9}}$ cm$^{-2}$, $\sim{10^{10}}$ cm$^{-2}$ and $\sim{10^{12}}$ cm$^{-2}$, respectively. The phonon-lines are marked by straight arrows and emission features by tilted cyan solid lines. White braces indicate the energy positions of IMBE. Particularly the m$_1 \rightarrow$ m$_2$ transition is enhanced with increasing doping and hence position of E$_F$  in agreement with jDOS(E$_F$) [T = 4K]. 
}
\end{figure*}

Excitation-power dependent RILS measurements are performed for the 3$^o$ tWSe$_2$ sample in order to vary E$_F$ by the photo-induced holes. RILS spectra from a different set of experiments are compared in Fig. \ref{fig:Fig.SI_RILS_P} for (a) 1$\mu$W corresponding to a change in the hole density by $\Delta$p $\thickapprox10^9$cm$^{-2}$, (b) 50$\mu$W ($\Delta$p $\thickapprox10^{10}$cm$^{-2}$) and (c) 1mW ($\Delta$p $\thickapprox10^{12}$cm$^{-2}$), respectively.\\

Due to the strong band non-parabolicity and the unknown intrinsic doping it is not possible to estimate the change in E$_F$ from $\delta$p. 
While it is not possible within the experimental certainty to clearly identify a shift in the energy, the intensity of the RILS mode assigned to IMBE m$_1 \rightarrow$m$_2$ is significantly reduced for the low density measurements as expected from the jDOS. Similarly, for the high energy transition that seems to be most pronounced for the higher power measurements indicated by an even broader overlapping emission band towards 100meV and beyond. This is most pronounced for 50$\mu$W. For the 1mW spectra some spectral weight shifts back to the intermediate transitions regions m$_1 \rightarrow$m$_{4,5}$. This also agrees with the jDOS from the band structure calculation and strongly supports our interpretation that the observed broad RILS signatures are indeed collective IMBE between moir\'e bands at the VBM.

\clearpage
\subsection*{Lineshape-analysis of power dependent RILS data  of tWSe$_2$ ($t=3°$)}
\begin{figure*}[!htb]
        \center{\includegraphics[width=\textwidth]
        {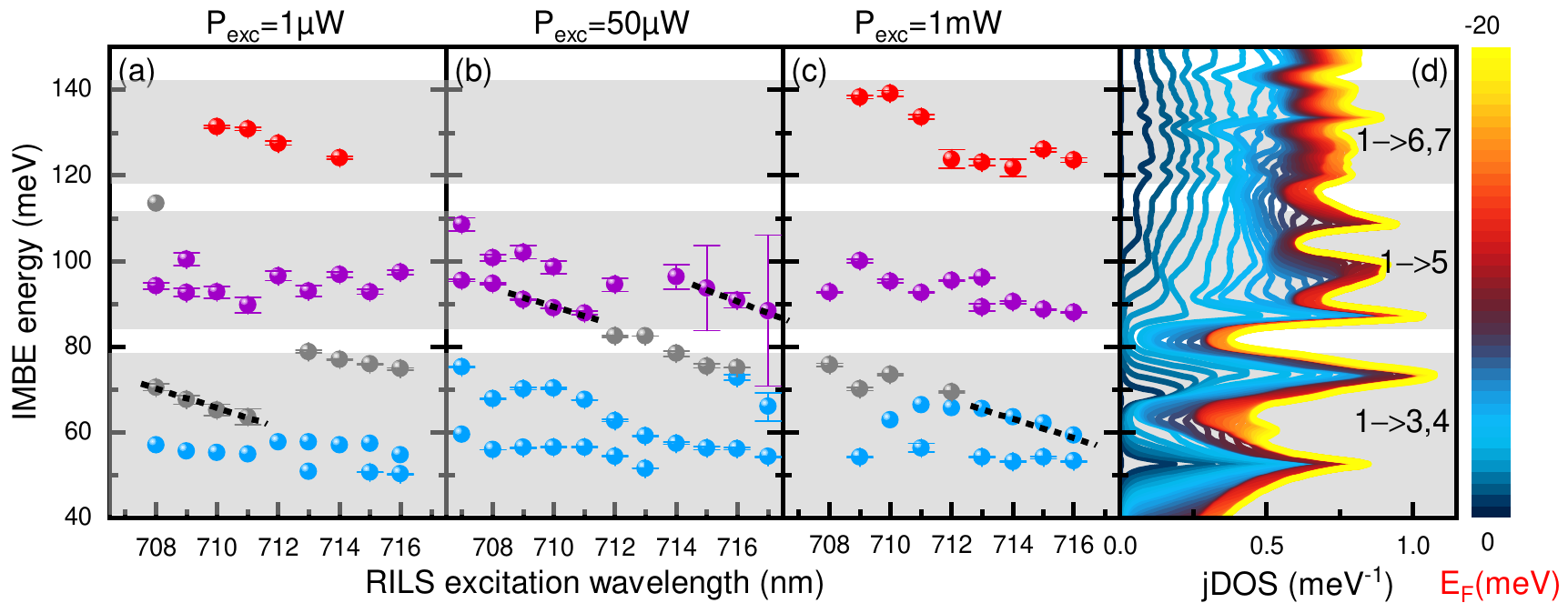}}
        \caption{\label{fig:anaylsis} 
 Extracted peak energies from a line-shape analysis to the excitation power P$_{Laser}$ dependent RILS measurements displayed in SI-Fig. \ref{fig:Fig.SI_RILS_P} taken on the 3$^o$ tWSe$_2$ bilayer at a temperature of T = 4K at three different excitation powers of (a) 1$\mu$W, (b) 50$\mu$W and (c) 1mW, respectively. The corresponding changes in the photo-generated hole-densities are $\sim{10^{9}}$ cm$^{-2}$, $\sim{10^{10}}$ cm$^{-2}$ and $\sim{10^{12}}$ cm$^{-2}$, respectively. The lineshape analysis was performed as described in the main text by numerically describing the spectra fitting a suitable sum of Lorentzian curves to the spectra in order to deconvolute the dominating contribution to the spectra. Most of the peak energies of the Lorentz curves are constant by changing the excitation wavelength unambiguously identifying them RILS on collective excitations. The dashed lines mark data points where the RILS signal is likely superimposed by emission and hence slightly descend by increasing excitation wavelength. The combined fit error is in the order of a few percent. (d) Calculated joint density of states jDOS for IMB transitions in dependence of the Fermi-level E$_F$  for 3$^o$ tWSe$_2$ for vertical transitions.}
\end{figure*}

\begin{figure*}[!htb]
        \center{\includegraphics[width=0.8
    \textwidth]
        {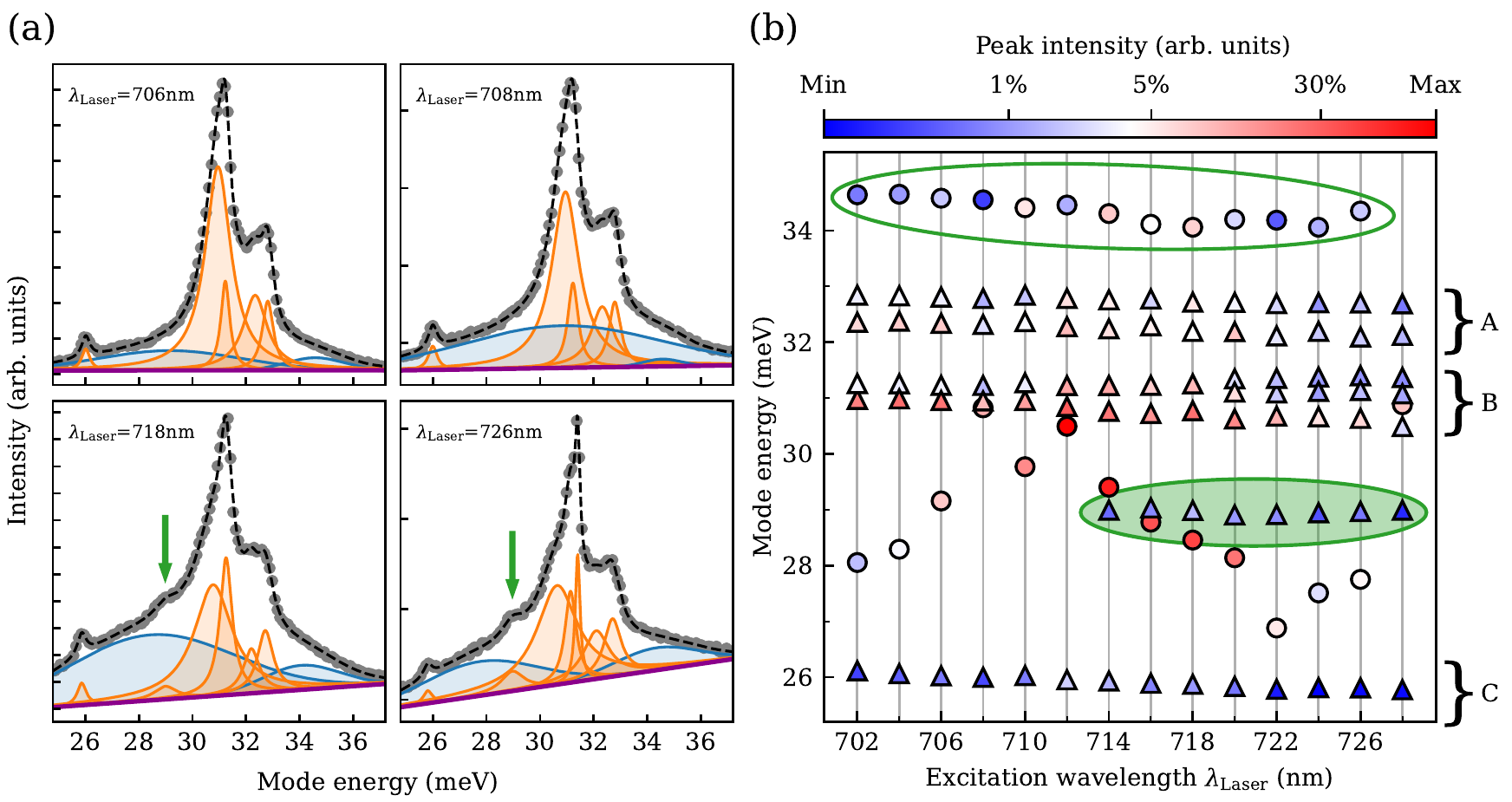}}
        \caption{\label{fig:Fig.SI_RILS_LA} 
(a) Example RILS spectra of the lower energy range between 24meV and 38meV (points) for the 3$^o$ tWSe$_2$ bilayer. The black dashed lines are the sum of a combination of Lorentz (orange) and Gauss (blue) curves in order to identify the individual contribution to the spectra after subtraction of a linear background (magenta). Overall, the described fit approach provides a very good numerical description of the experimental RILS spectra. In contrast to the higher energy range, we find a set of rather sharp modes that can be described by Lorentz terms in addition to one or two rather broad modes well described by Gaussian terms. The latter change with excitation wavelength and have similar absolute peak energies and are therefore assigned to PL. The relative energy shifts of the Lorentz curves are independent from the excitation wavelengths and are therefore assigned to inelastic scattering on collective excitations. Most of the Lorentz contributions are assigned to first order (Raman active) and second order Raman processes (enhanced by resonant excitation) by scattering on phonons in WSe$_2$. One mode at 28.8 meV cannot be assigned to phonon contributions and occurs only under extreme resonance. This mode is assigned to a collective IMBE between m$_1$ and m$_2$ moir\'e miniband. (b) Summary of the peak position as a function excitation wavelengths. The individual peak intensities are color coded using a logarithmic scale demonstrating the resonance conditions. Triangular data points are from Lorentzian fit curves assigned to inelastically scattered light due to their constant energy shifts and circular data  are from Gaussian fit curves and are assigned to superimposed emission most likely due to defect-PL or in case of the signal being rather stable in energy for different excitation wavelength that might be also due to collective electronic excitations. The peak ensembles marked by A, B and C are interpreted as phonons. The RILS mode occurring at an energy of 28.8 meV only under extreme resonance in several set of data (not shown) for an excitation wavelength range for 714nm $\leq \lambda_{Laser} \leq$ 728nm is interpreted as a collective electronic excitation. This mode is tentatively interpreted as IMBE between first and second moir\'e miniband m$_1 \rightarrow$ m$_2$ in agreement with theory, but we cannot exclude an alternative interpretation as collective electronic excitation between the spin-split conduction bands CB$_- \rightarrow$ CB$_+$ at the K points [see Fig.2 (a) in the main text]. For the assignment of the phonon modes we follow the combined experimental and theoretical work by Liam P. McDonnell and colleagues \cite{McDonnell_2020} as follows. Region A: E$_{LO}^{'}$(K), A$_{1}^{'}$(K) or 2LA(M);  Region B: E$_{LO/TO}^{'}$($\Gamma$), A$_{1}^{'}$($\Gamma$), LA(M)+ZA(M); Region C (dispersing mode): 2TA(K) or TA(M)+ZA(M); }
\end{figure*}

\clearpage

\subsection*{Temperature and polarization dependent RILS measurements on tWSe$_2$ ($t=8°$)}

\begin{figure*}[!htb]
        \center{\includegraphics[width=0.9\textwidth]
        {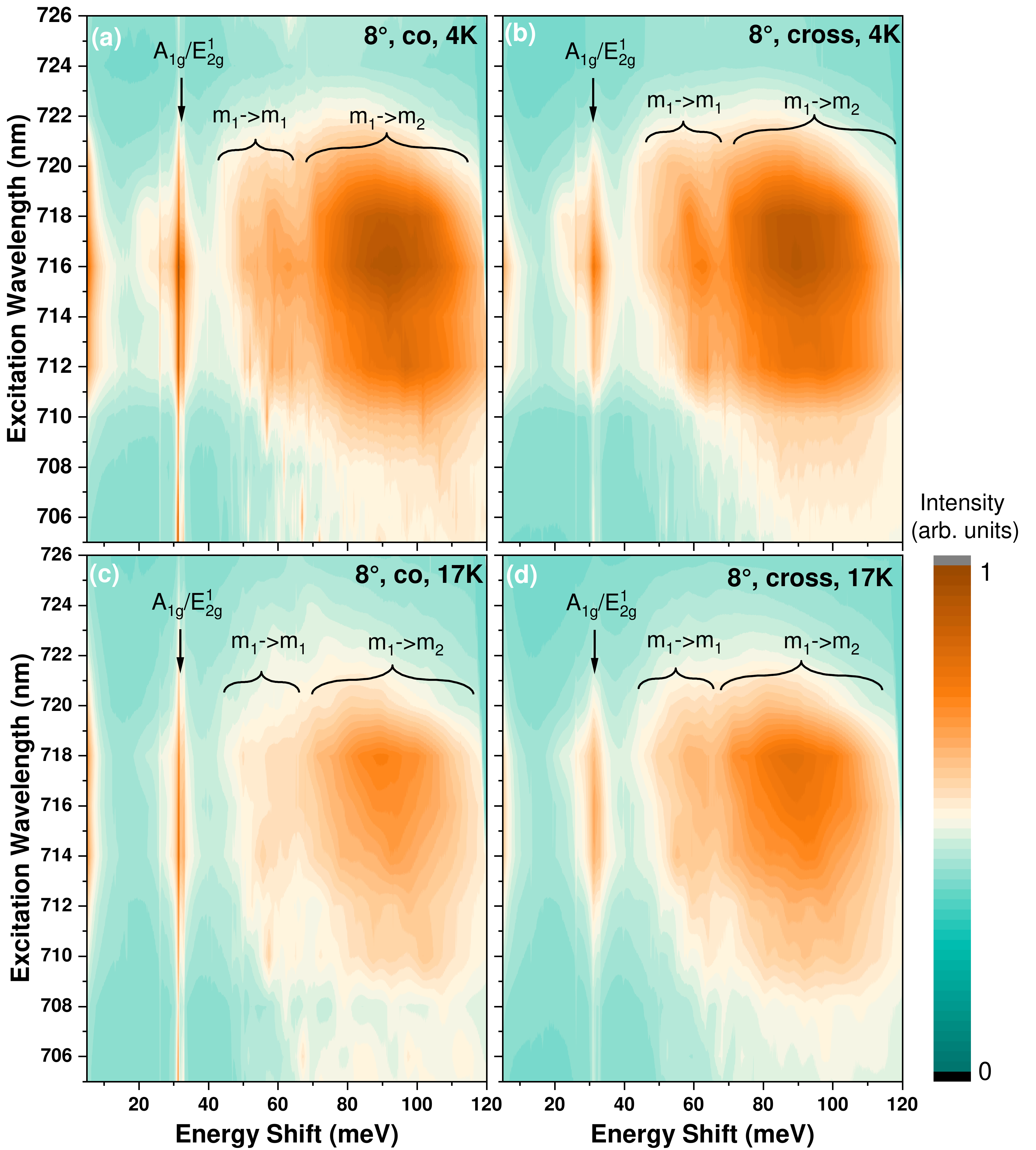}}
        \caption{\label{fig:Fig.SI_RILS_8} 
False color representation of RILS spectra for 8$^o$ tWSe$_2$. Linear co- and cross scattering geometries are compared for two different temperatures. (a) co-polarized RILS spectra at T = 4K; (b) cross-polarized RILS spectra at T = 4K; (c) co-polarized RILS spectra at T = 17K; (d) cross-polarized RILS spectra at T = 17K  [P = 500$\mu$W]. The phonon modes and collective charge density excitations due to intra moir\'e band transitions m$_1 \rightarrow$ m$_1$ as well as inter moir\'e band transitions m$_1 \rightarrow$ m$_2$ are indicated. While we do not observe significant polarization dependence for the collective charge density exciation, those modes are smeared out by increasing the temperature from 4K to 17K.}
\end{figure*}

\clearpage
\subsection*{Calculated mini-band structure}

\begin{figure*}[!htb]
        \center{\includegraphics[width=0.7\textwidth]
        {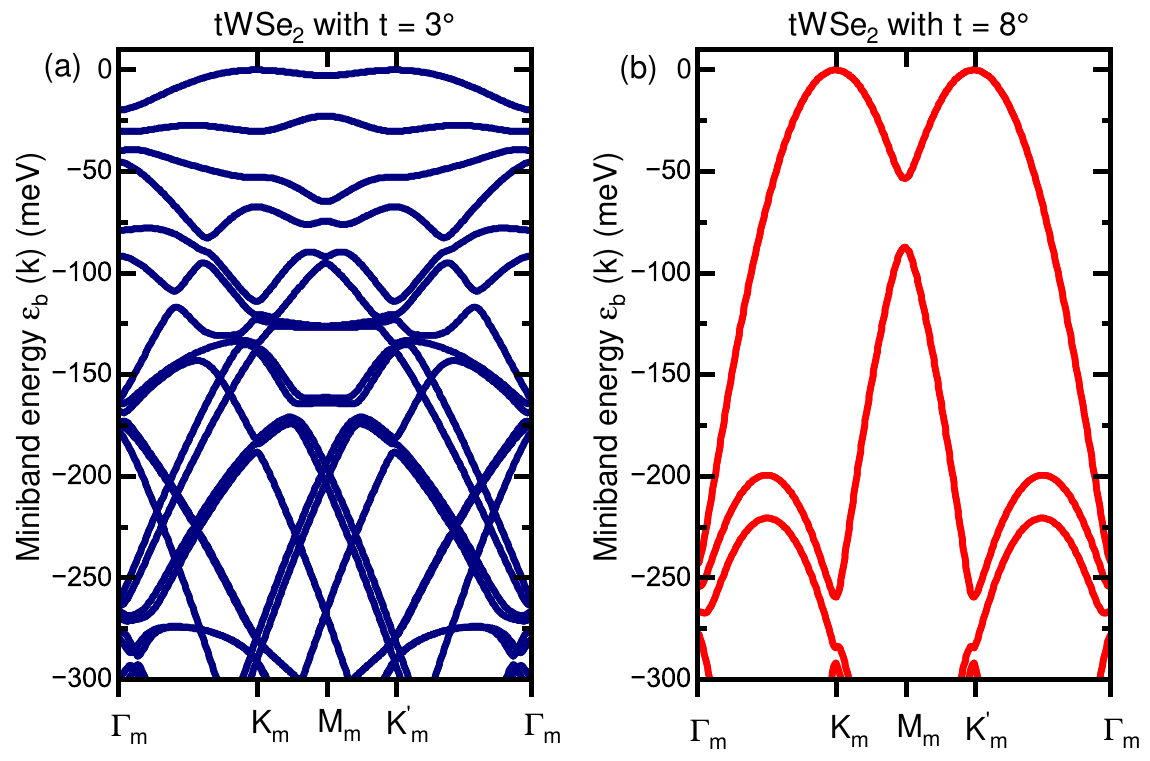}}
        \caption{\label{fig:Fig.SI_minibands} 
 Calculated energy dispersion of the highest moir\'e mini bands down to -300meV around the K, K' states for a 3$^o$ tWSe$_2$ bilayer (a) and a 8$^o$ tWSe$_2$ bilayer (b). 
}
\end{figure*}

\clearpage
\subsection*{Calculated jDOS for tWSe$_2$ ($t=3°\pm 0.3°$) }

\begin{figure*}[!htb]
    \centering
    \includegraphics{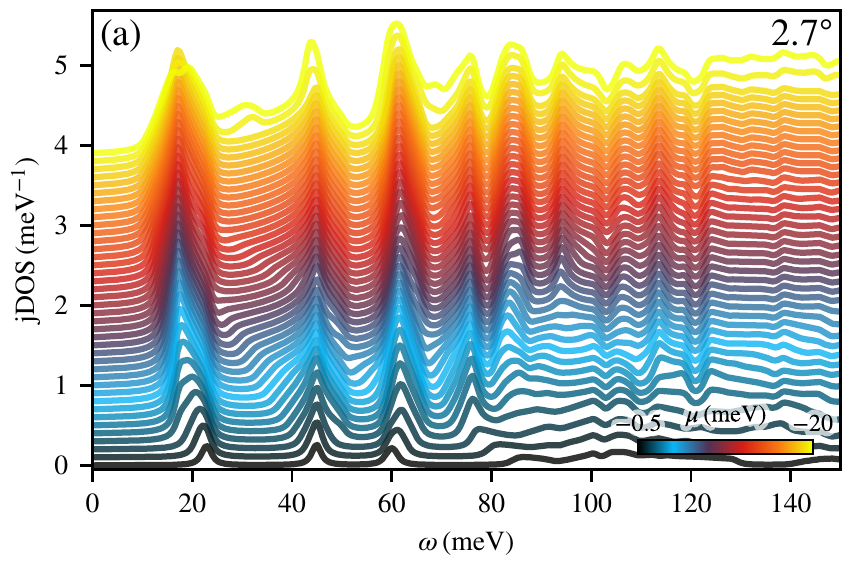} \\
    \includegraphics{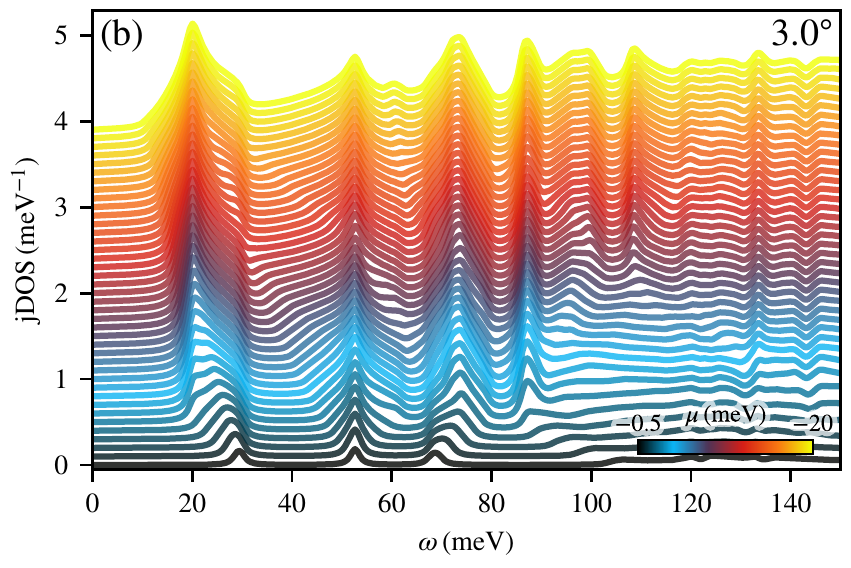} \\
    \includegraphics{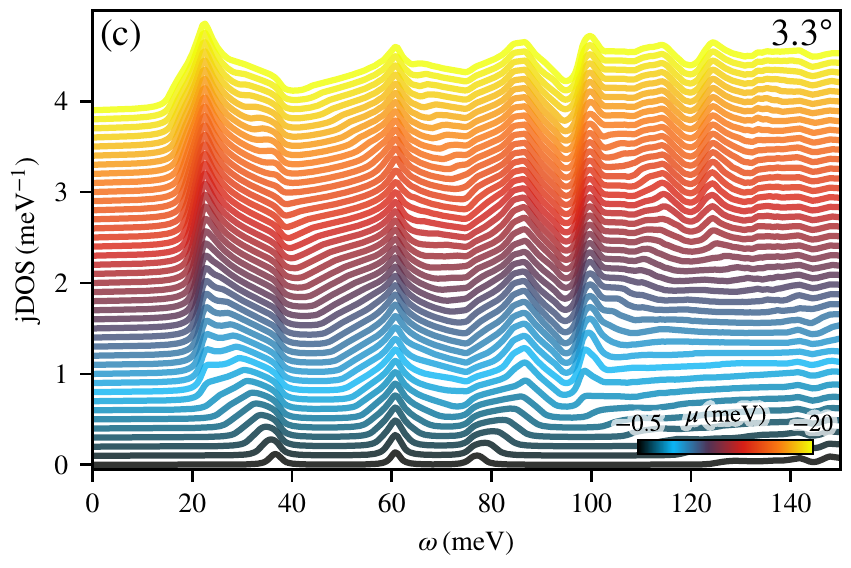}
    \caption{\label{fig:Fig.jDOS} 
 Calculated jDOS from the energy dispersion of the highest moir\'e mini bands  the K, K' states for (a) 2.7$^o$ tWSe$_2$ bilayer (b) 3.0$^o$ tWSe$_2$ bilayer and (c) 3.3$^o$ tWSe$_2$ bilayer. The curves are offset by $0.1\,\mathrm{meV}^{-1}$ for clarity.
}
\end{figure*}
\clearpage

%

\end{document}